\documentclass[aps,pra,twocolumn,superscriptaddress,showpacs]{revtex4}
\def\be{\begin{equation}}
\def\ee{\end{equation}}
\def\bea{\begin{eqnarray}}
\def\eea{\end{eqnarray}}
\def\bi{\begin{itemize}}
\def\ei{\end{itemize}}

\usepackage{graphicx}
\usepackage{amsmath}
\usepackage{amssymb}

\begin{document}

\title{ Dynamics of a quantum phase transition with decoherence:  \\
          the quantum Ising chain in a static spin environment       }

\author{Lukasz Cincio}
\author{Jacek Dziarmaga}
\author{Jakub Meisner}
\author{Marek M. Rams}
\affiliation{ Institute of Physics and 
              Centre for Complex Systems Research, 
              Jagiellonian University,
              Reymonta 4, 30-059 Krak\'ow, 
              Poland}

\begin{abstract}
We consider a linear quench from the paramagnetic to ferromagnetic phase
in the quantum Ising chain interacting with a static spin environment. 
Both decoherence from the environment and non-adiabaticity of the evolution 
near a critical point excite the system from the final ferromagnetic 
ground state. For weak decoherence and relatively fast quenches the excitation 
energy, proportional to the number of kinks in the final state, decays like 
an inverse square root of a quench time, but slow transitions or strong 
decoherence make it decay in a much slower logarithmic way. We also find that 
fidelity between the final ferromagnetic ground state and a final state after 
a quench decays exponentially with a size of a chain, with a decay rate proportional 
to average density of excited kinks, and a proportionality factor evolving from 
$1.3$ for weak decoherence and fast quenches to approximately $1$ for slow 
transitions or strong decoherence. Simultaneously, correlations between kinks 
randomly distributed along the chain evolve from a near-crystalline anti-bunching 
to a Poissonian distribution of kinks in a number of isolated Anderson localization 
centers randomly scattered along the chain. 
\end{abstract}

\maketitle

\section{Introduction}

Phase transition is a fundamental change in the state of a system when one 
of its parameters passes through the critical point. In a second order phase 
transition, the fundamental change is continuous and the critical point is 
characterized by divergences in the correlation length and in the relaxation 
time. This critical slowing down implies that no matter how slowly a system 
is driven through the transition, its evolution cannot be adiabatic close to 
the critical point. As a result, ordering of the state after the transition 
is not perfect: it is a mosaic of ordered domains whose finite size $\hat\xi$ 
depends on the rate of the transition. This scenario was first described in 
the cosmological context by Kibble \cite{K} who appealed to relativistic 
casuality to set the size of the domains. The dynamical mechanism relevant for 
second order phase transitions was proposed by Zurek \cite{Z}. It is based on 
the universality of critical slowing down, and leads to a prediction that average 
size $\hat\xi$ of the ordered domains scales with the transition time $\tau_Q$ as 
\be
\hat\xi~\simeq~\tau_Q^{\frac{\nu}{z\nu+1}}~,
\label{xiKZ}
\ee
where $\nu$ and $z$ are critical exponents. The Kibble-Zurek mechanism (KZM) for 
second order thermodynamic phase transitions was confirmed by numerical simulations of 
the time-dependent Ginzburg-Landau model \cite{KZnum} and successfully tested by 
experiments in liquid crystals \cite{LC}, superfluid helium 3 \cite{He3}, both 
high-$T_c$ \cite{highTc} and low-$T_c$ \cite{lowTc} superconductors, and even in
non-equilibrium systems \cite{ne}. With the exception of superfluid $^4$He -- 
where the early  detection of copious defect formation \cite{He4a} was 
subsequently attributed to vorticity inadvertently introduced by stirring \cite{He4b}, 
and the situation remains unclear -- experimental results are consistent with KZM, 
although more experimental work is clearly needed to allow for more stringent 
experimental tests of KZM. Quite recently, a new experiment was reported in 
Ref. \cite{Anderson} where they observe, for the first time, spontaneous appearance
of vorticity during Bose-Einstein condensation driven by evaporative cooling confirming
KZM predictions in Ref. \cite{BEC}.

The Kibble-Zurek mechanism is thus a universal theory of the dynamics of second
order phase transitions whose applications range from the low temperature
Bose-Einstein condensation (BEC) to the ultra high temperature transitions in the 
grand unified theories of high energy physics. However, the zero temperature quantum 
limit remained unexplored until very recently and quantum phase transitions are in 
many respects qualitatively different from transitions at finite temperature. Most 
importantly time evolution is unitary, so there is no damping, and there are no thermal 
fluctuations that initiate symmetry breaking in KZM. The recent progress on dynamical 
quantum phase transitions is mostly theoretical, see Refs. \cite{3sites,Bodzio1,KZIsing,
Dziarmaga2005,Polkovnikov,Cucchietti,Levitov,Cincio,Bodzioferro,Hindusi,GLM,Kitaev,
Meisner,Schaller,Polkovnikov_optimal} and, for an example of a disordered quantum system, 
Ref. \cite{JDrandom}, but there is already one significant exception: the experiment in 
Ref. \cite{ferro} on a transition from paramagnetic to ferromagnetic phase in a dipolar BEC. 
Generic outcome of that experiment is a mosaic of finite-size ferromagnetic domains whose origin was 
attributed to the Kibble-Zurek mechanism. This explanation is further supported by theory 
in Ref. \cite{Bodzioferro}. 

A majority of theoretical work was devoted to the prototypical exactly solvable quantum 
Ising chain \cite{KZIsing,Dziarmaga2005,Levitov,Cincio,Fubini,Schaller,Polkovnikov_optimal},  
\be
H_S~=~-\sum_{n=1}^N \left[ g~\sigma^x_n + \sigma^z_n\sigma^z_{n+1} \right]~
\label{HS}
\ee 
driven by a linear quench 
\be
g(t)~=~-\frac{t}{\tau_Q}~
\label{gt}
\ee  
from $g=\infty$ to $g=0$ i.e. across the quantum phase transition from paramagnet
to ferromagnet at $g_c=1$. Since the system is gapless at $g_c$, when $\tau_Q\gg1$ 
the evolution becomes non-adiabatic at
\be
\hat g-g_c~\simeq~\tau_Q^\frac{1}{z\nu+1}~,
\label{hatg}
\ee
see Ref. \cite{KZIsing}. The freezeout at $\hat g$ is the closer to $g_c$ the 
slower is the transition. The Ising critical exponents $\nu=1$ and $z=1$ determine 
an average size of ferromagnetic domains at $g=0$ proportional to the correlation length at 
$\hat g$:
\be
\hat\xi\simeq\tau_Q^{1/2}~,
\label{xi12}
\ee
or final density of kinks (domain walls) at $g=0$
\be
d ~\simeq~ \hat\xi^{-1} ~\simeq~ \tau_Q^{-1/2}~
\label{dpure}
\ee
which is proportional to excitation energy density. Other properties, like 
spin-spin correlation functions \cite{Cincio,Levitov} or entropy of entanglement 
between a block of consecutive spins and the rest of the chain \cite{Cincio} were 
analyzed in some detail and they all turned out to be determined by $\hat\xi$. 

While the quench in the isolated system (\ref{HS}) seems to be well analyzed, 
relatively little is still known about dynamical transitions in open quantum systems 
subject to interaction with environment. Significant progress was made in 
Ref. \cite{Fubini} in a ``global'' case when interaction between the system $S$ with 
the Hamiltonian (\ref{HS}) and its environment $E$ is described by    
$V=R\left(\sum_n \sigma^x_n\right)$, where $R$ is a hermitian operator of
the environment. This global model is solvable thanks to its translational invariance
and its solutions indicate that decoherence is increasing density of excited kinks
as compared to an isolated system. A local model, with the system (\ref{HS})
coupled to an {\it Ohmic} heat bath, was analyzed in Ref. \cite{Faziodecoherence}
distinguishing between different regimes of parameters where defect production is 
dominated either by KZM or external heating. In this paper, we propose a quite realistic, 
but still solvable, model of local zero temperature decoherence from a {\it static}
environment. Its solution predicts dramatic increase in the density $d$ of excited 
kinks as compared to an isolated system with $d$ decaying as only a logarithmic 
function of $\tau_Q$.

Motivation for this study is twofold. It comes both from condensed matter physics,
where it is virtually impossible to isolate a system from its environment, and adiabatic 
quantum computation, where a system is initially prepared in a simple ground state of 
a simple initial Hamiltonian $H_0$ and then it is driven adiabatically to a final 
Hamiltonian $H_1$ whose non-trivial ground state is the desired solution of a complex 
computational problem. The computation is complicated by a quantum critical point
somewhere on the way from $H_0$ to $H_1$ which can make the adiabaticity problematic,
but see Refs. \cite{Schaller,Polkovnikov_optimal} for methods how to circumvent this
problem. The Ising chain (\ref{HS}) is a toy model of adiabatic quantum computer with the 
final (trivial) ferromagnetic ground state at $g=0$ playing the role of the desired 
``non-trivial'' ground state. When the ``computer'' is isolated from environment, then 
Eq. (\ref{dpure}) implies that the minimal ``computation time'' $\tau_Q$ required to keep 
the evolution adiabatic or, equivalently, to make $\hat\xi\gg N$ is
\be
\tau_Q^{\rm isolated}~\simeq~N^2~. 
\label{poly}
\ee      
The ``isolated'' computation problem is polynomial in $N$. In contrast, in our model
of decoherence similar argument predicts
\be
\tau_Q^{\rm open}~\simeq~e^{\sqrt{N}}
\label{tauQopen}
\ee
which is non-polynomial in $N$.

\section{Ising chain in static spin bath}

In this paper we couple the Ising chain (\ref{HS}) to an environment $E$ of 
$M$ spins through the interaction 
\be
V~=~-\sum_{n=1}^N \sum_{m=1}^M \sigma^x_n ~V_{nm}~\tau^x_m~.
\ee
Here $\tau_m$'s are Pauli matrices of environmental spins. The spins 
are static, with $H_E=0$, and the total Hamiltonian is just $H=H_S+V$.

Initially at $t\to-\infty$ the system is in the ground state $|0_{g\to\infty}\rangle$
of the pure Ising chain (\ref{HS}) with all spins polarized along $x$. This assumption
is self-consistent in our open system because large initial energy gap of $2g$ 
makes the influence of the static environment so negligible that the initial states of 
$S$ and $E$ can be assumed uncorrelated: $\rho_{S+E}=\rho_S\otimes\rho_E$ with 
$\rho_S=|0_{g\to\infty}\rangle\langle0_{g\to\infty}|$ and the environment is initially 
in a pure state 
\be
\sum_{s_1,...,s_M=-1,+1}
c_{s_1,...,s_M}~
|s_1\rangle ... |s_M\rangle~.
\ee
Here $\tau^x_m|s_m\rangle=s_m|s_m\rangle$.

After evolution for time $\Delta t$ reduced density matrix of the system 
$\rho_S={\rm Tr}_E\rho_{S+E}$ becomes
\bea
\rho_S(\Delta t) &=&
\sum_{\vec s} 
\left|c_{\vec s}\right|^2~ 
U(\Delta t,\vec s)~
|0_\infty\rangle\langle0_\infty|~
U^\dag(\Delta t,\vec s) ~ 
\nonumber\\
&\equiv&
\overline{
U(\Delta t,\vec s)~
|0_\infty\rangle\langle0_\infty|~
U^\dag(\Delta t,\vec s)
}~.
\label{rhoS}
\eea
Here $\vec s=(s_1,...,s_M)$,
\be
U(\Delta t,\vec s)~=~
{\cal T}~
\exp{\left[-i\int_0^{\Delta t} dt'~H(t',\vec s)\right]}~, 
\ee
and 
\be
H(t,\vec s)~=~
-\sum_{n=1}^N 
\left\{ 
\left[g(t)+\Gamma_n\right]~\sigma^x_n + \sigma^z_n\sigma^z_{n+1} 
\right\}~
\label{Hsigma}
\ee 
with random magnetic fields of the static environment
\be
\Gamma_n~=~\sum_{m=1}^M V_{nm} s_m~.
\label{gamman}
\ee
The overline in Eq. (\ref{rhoS}) is an average over $\vec s$ with probability
distribution $|c_{\vec s}|^2$, but it can also be interpreted as an average 
over random ``disorder'' field $\Gamma_n$. $\rho_S(\Delta t)$ is an average over 
states $U(\Delta t,\vec s)|0_\infty\rangle$ obtained in quenches with different 
disordered Hamiltonians (\ref{Hsigma}). In this way, our original problem of a 
quench in the open pure Ising model (\ref{HS}) is mapped to an average over 
quenches in the isolated random Ising model (\ref{Hsigma}).

In the following, rather than struggle with the problem in its full generality,
we assume that each spin of the environment couples to only one spin of the system
or, in other words, each spin of the system has its own {\it local} environment.
Consequently, $\Gamma_m$ and $\Gamma_n$ are statistically independent when $m\neq n$. 
We also assume that each $\Gamma_n$ has the same Gaussian probability distribution 
\be
f(\Gamma)~=~\frac{e^{-\Gamma^2/2\sigma^2}}{\sqrt{2\pi\sigma^2}}~,
\label{gauss}
\ee
where $\sigma$ is strength of disorder/decoherence.

The Hamiltonian (\ref{Hsigma}) belongs to the universality class of the well 
known random Ising chain \cite{random,DFisher}. It has a continuous quantum 
critical point at $g_c$ when 
\be
\overline{\ln|g_c+\Gamma|}~\equiv~
\int_{-\infty}^{\infty} d\Gamma~f(\Gamma)~\ln|g_c+\Gamma|~=~
0~.
\label{gc}
\ee
$g_c$ depends on $\sigma$ as shown in Fig. \ref{FigStatic}A. There is no critical point 
for $\sigma>1.887$ when the disorder is too strong. No matter how weak $\sigma$ is, 
renormalization group transformations drive the model (\ref{Hsigma}) towards an infinite 
disorder fixed point with different critical exponents than in the pure Ising chain: 
the random chain has $\nu=2$ instead of $\nu=1$ and $z\to\infty$ instead of $z=1$ \cite{DFisher}. 
A straightforward application of the standard KZ formula (\ref{xiKZ}) gives 
$\hat\xi\simeq 1$ i.e. a domain size that does not depend on the quench time $\tau_Q$. 
A more careful argument in Ref. \cite{JDrandom}, going back to the basics of KZM, predicts 
a logarithmic dependence 
\be
\hat\xi~\simeq~
\frac{\ln^2\left[\alpha\tau_Q\right]}
     {\ln^2\left[\ln(\alpha\tau_Q)\right]}~,
\label{xirandom}
\ee
with a non-universal $\alpha\simeq 1$. This equation leads to the estimate in 
Eq. (\ref{tauQopen}). Since Eq. (\ref{xirandom}) is based on the universality class 
alone, it is valid for any model of this class when $\tau_Q$ is long enough for 
the quench to become non-adiabatic close enough to $g_c$ to be affected by disorder. 
The estimate (\ref{xirandom}) was confirmed by numerics in the model of Ref. \cite{JDrandom},
where it was also found that for weak disorder and relatively fast $\tau_Q$ one 
recovers Eqs. (\ref{xi12},\ref{dpure}) as in the pure model Eq. (\ref{HS}). 

In the present effective model (\ref{Hsigma}), it is simple to estimate how slow a quench
needs to be for Eqs. (\ref{xi12},\ref{dpure}) and, more importantly, Eq. (\ref{poly}) 
to be not valid. Assuming that influence of $\Gamma_n$ is negligible, evolution 
becomes non-adiabatic at a field $\hat g$ in Eq. (\ref{hatg}). This assumption 
is not self-consistent when the remaining distance from $\hat g$ to $g_c$, 
$\hat g-g_c\simeq\tau_Q^{-1/2}$, is less than the strength $\sigma$ of disorder field
$\Gamma_n$, or equivalently
\be
\tau_Q~\sigma^2~\gg~1~.
\label{strong}
\ee
Thus, no matter how weak the decoherence is, its influence is not negligible
when the transition is slow enough: $\tau_Q\gg\sigma^{-2}$. In consequence,
there is a maximal number of qubits
\be
N~\ll~\frac{1}{\sigma}~
\label{Nmax}
\ee
which can be simulated with polynomial efficiency, compare Eq. (\ref{poly})
and (\ref{strong}).

In the next Section, we review static properties of the random Ising model 
(\ref{Hsigma}) many of which, we believe, are described in this form for the 
first time.

\section{Random Ising chain}

Here we assume for convenience that $N$ is even and following Refs. \cite{LSM,JDrandom} 
make the Jordan-Wigner transformation $\sigma^x_n=1-2 c^\dagger_n c_n$ and 
$\sigma^z_n=-\left( c_n+ c_n^\dagger\right)\prod_{m<n}(1-2 c^\dagger_m c_m)$
introducing spinless fermionic operators $c_n$. The Hamiltonian (\ref{Hsigma}) becomes 
$H=P^+H^+P^++P^-H^-P^-$ where
$P^{\pm}=\frac12\left[1~\pm~\prod_{n=1}^N\left(1-2c_n^\dagger c_n\right)\right]$ are 
projectors on subspaces with even ($+$) and odd ($-$) numbers of $c$-quasiparticles and  
\bea
H^{\pm}&=&
\sum_{n=1}^N
\left(
g_n c_n^\dagger  c_n - c_n^\dagger  c_{n+1} - c_{n+1}  c_n - \frac{g_n}{2} 
\right)~ 
\nonumber\\
&+&
~{\rm h.c.}~
\label{Hpm}
\eea
are quadratic Hamiltonians. Here 
\be
g_n=g+\Gamma_n
\label{gn}
\ee 
for short. The $c_n$'s in $H^-$ satisfy periodic boundary conditions $c_{N+1}=c_1$, 
but the $c_n$'s in $H^+$ must obey $c_{N+1}=-c_1$, what we call ``antiperiodic'' 
boundary conditions. 

The parity of the number of $c$-quasiparticles is a good quantum number and the 
ground state has even parity for any value of $g$. Assuming that the quench begins 
in the ground state we can confine to the subspace of even parity. In this subspace 
the quadratic $H^+$ is diagonalized by a Bogoliubov transformation
$c_n=\sum_{m=1}^{N}(u_{nm}\gamma_m + v^*_{nm} \gamma_m^\dagger)$. 
The index $m$ numerates (Bogoliubov) eigenmodes of the stationary Bogoliubov-de Gennes equations
\bea
\omega_m\frac{du_{n,m}^\pm}{dt}=2 g_n u_{n,m}^\mp - 2 u_{n-1,m}^\mp ~
\label{BdG}
\eea
with $\omega_m>0$.
Here we define $u_{nm}^\pm\equiv u_{nm} \pm v_{nm}$ and assume the anti-periodic
boundary conditions: $u^\pm_{N+1,m}=-u^\pm_{1,m},u^\pm_{0,m}=-u^\pm_{N,m}$.
The eigenstates $(u_{nm},v_{nm})$, normalized so that $\sum_n\left(|u_{nm}|^2+|v_{nm}|^2\right)=1$, 
define quasiparticle operators $\gamma_m=u_{nm}^*c_n+v_{nm}^*c_n^\dagger$. 
After the Bogoliubov transformation the Hamiltonian becomes
$H^+=\sum_{m=1}^N\omega_m(\gamma_m^\dagger \gamma_m-\frac12)$
which is a simple-looking sum of quasiparticles. However, thanks to the projection 
$P^+~H^+~P^+$ only states with even numbers of quasiparticles 
belong to the spectrum of $H$.

\begin{figure}
\includegraphics[width=0.999\columnwidth,clip=true]{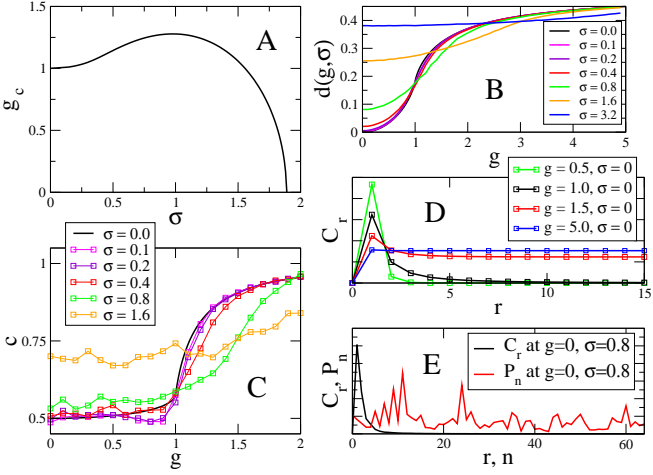}
\caption{ 
In A the critical $g_c$ in Eq. (\ref{gc}) is shown as a function of $\sigma$ in 
Eq. (\ref{gauss}). In B density of kinks in a ground state of the random chain
Eq. (\ref{Hsigma}) is shown as a function of $g$ for $N=512$ and different 
$\sigma$'s. In C we show correlation coefficient $c$ in Eq. (\ref{Fexp}) 
as a function of $g$ and $\sigma$. When $\sigma<1$, $c\approx 0.5$ in the
ferromagnetic phase below $g_c\approx 1$ and $c>0.5$ in the paramgnetic
phase above $g_c$. In D we show correlator $C_r$ in Eq. (\ref{C_r}) between 
two kinks in a Cooper pair in the pure Ising chain ($\sigma=0$). $C_r$ is 
localized below $g_c=1$ and delocalized above. In E both $C_r$ and $P_n$ in 
Eq. (\ref{P_n}) are shown in the ferromagnetic phase at $g=0$ and $\sigma=0.8$. 
Here $N=512$ in both D and E.
}
\label{FigStatic}
\end{figure}

When $g=\sigma=0$ the final system Hamiltonian (\ref{HS}) has $N$ degenerate 
quasiparticles with $\omega=2$. We choose an orthonormal basis 
\bea
u^{(0)}_{nm} &=& \frac12\left(\delta_{n+1,m}-\delta_{n,m}\right)~,\\
v^{(0)}_{nm} &=& \frac12\left(\delta_{n+1,m}+\delta_{n,m}\right)~,
\eea
antiperiodic in $n$. These eigenmodes define Bogoliubov quasiparticles 
$\gamma^{(0)}_m$ which are simply kinks localized at bonds between sites 
$m$ and $m+1$. The kinks are related to quasiparticles $(u_{nm},v_{nm})$ 
at finite $(g,\Gamma)$ by a Bogoliubov transformation
\be
\gamma_a~=~
U_{ba}^*~\gamma^{(0)}_{b}~+~
V_{ba}^*~\gamma^{(0)\dag}_{b}~,
\label{UV}
\ee
where 
\bea
U_{ba}&=&
\sum_n
\left( u_{nb}^{(0)*}u_{na}~+~v_{nb}^{(0)*}v_{na} \right)~,\\
V_{ba}&=&
\sum_n
\left( v_{nb}^{(0)}u_{na}~+~u_{nb}^{(0)}v_{na} \right)~
\eea
leading, for example, to a simple expression for average kink density 
in the ground state $|0_{g,\Gamma}\rangle$ of the Hamiltonian (\ref{Hsigma}) 
\be
d(g,\sigma)~=~
\frac{1}{N}
\overline{ 
\langle 0_{g,\Gamma}|
\sum_{m=1}^N \gamma_m^{(0)\dag}\gamma_m^{(0)}
|0_{g,\Gamma} \rangle
}~=~
\frac{{\rm Tr} V^\dag V}{N}~,
\label{dVV}
\ee
shown in Fig. \ref{FigStatic}B. It is finite at any $g>0$, both in the 
ferromagnetic $(g<g_c)$ and paramagnetic phase $(g>g_c)$, but, as we will 
see, kink-kink correlations are qualitatively different in the two phases.

These correlations can be indirectly probed by average fidelity $F$ between 
the final ground state $|0\rangle$ of $H_S$ at $g=0$ (state with no kinks 
$\gamma^{(0)}$) and ground states $|g,\Gamma_n\rangle$ at finite $g$ or $\sigma$:
\be
|0_{g,\Gamma}\rangle~=~
{\cal N}~
e^{\frac12\sum_{a,b=1}^N Z_{ab}\gamma_a^{(0)\dag}\gamma_b^{(0)\dag} }~
|0\rangle~.
\label{expZ}
\ee
Here 
$
Z~=~V^*(U^*)^{-1}
$ 
and ${\cal N}$ is a normalization factor. The average fidelity is
\bea
F&=&
\overline{\langle 0|0_{g,\Gamma}\rangle\langle 0_{g,\Gamma}|0\rangle}
\nonumber\\
&=&
\overline{1/
          \left(
          \sum_{n=0}^{N/2}
          \frac{1}{(n!)^2}
          \langle 0|\left(\hat Z^\dag\right)^n \hat Z^n|0\rangle
          \right)
} 
\nonumber\\
&=&
\overline{
{\rm Det}\left(1+Z^\dag Z\right)^{-1/2}
}
~.
\label{expTrlog}
\eea
Here $\hat Z\equiv\frac12\sum_{ab}Z_{ab}\gamma_a^{(0)\dag}\gamma_b^{(0)\dag}$ for short. 

We found that the fidelity is exponential in $N$
\be
F(g,\sigma,N)~\sim~(~1~-~c~d~)^N
\label{Fexp}
\ee
when $F\ll 1$. Here $d(g,\sigma)$ is the density of kinks in Eq. (\ref{dVV}) and 
Fig. \ref{FigStatic}B. The coefficient $c(g,\sigma)$ is shown in Fig. \ref{FigStatic}C. 
For weak disorder, when $\sigma\ll 1$, we have $c\approx \frac12$ when $g<g_c$ with 
$c$ increasing towards $1$ when $g\gg g_c$ or $\sigma\gg 1$. These two limits can be 
explained as follows.

When the magnetic fields $g_n=\Gamma_n+g$ in the effective Hamiltonian 
(\ref{Hsigma}) are strong, because either $g$ or $\sigma$ or both are strong, 
then in any ground state $|0_{g,\Gamma}\rangle$ all spins are polarized along 
$\sigma^x$. Fidelity to the $\sigma^z$-ferromagnet is $F=1/2^N$ and density
of kinks is $d=1/2$, hence $F=(1-cd)^N$ with $c=1$. This $c$ is consistent with 
the data shown in Fig. \ref{FigStatic}C when $g$ or $\sigma$ are strong.

In the opposite limit of weak magnetic fields $|g_n|=|g+\Gamma_n|\ll1$, the 
ground states are
\be
|0_{g,\Gamma}\rangle~\approx~
\prod_{n=1}^N
\frac{|\uparrow_n\rangle\pm\frac{g_n}{4}|\downarrow_n\rangle}
     {\sqrt{1+\frac{g_n^2}{16}}}~.
\label{0pert}
\ee
Their fidelity to the $\sigma^z$-ferromagnet is
\be
F~=~
\prod_{n=1}^N \frac{1}{1+\frac{g_n^2}{16}}~\approx~
\left(1-\frac12d\right)^N
\ee
when $F\ll1$. Here $d=\frac18\overline{g_n^2}\ll1$ is small density of kinks and 
$c=1/2$. This $c$ is consistent with the data in Fig. \ref{FigStatic}C when 
$\sigma\ll 1$ and $g<g_c\approx 1$.

It is interesting to interpret the widely different values of $\frac12\leq c<1$ in 
terms of a simple Poissonian model where each of $N$ bonds is either excited (with 
probability $d_{\rm exc}$) or not excited (with probability $1-d_{\rm exc}$) 
{\it independently} of other bonds. Here $d_{\rm exc}$ is average density of 
excitations. The fidelity is a probability that none of the $N$ independent bonds 
is excited 
\be
F~=~(~1~-~d_{\rm exc}~)^N~.
\label{Fmodel}
\ee 
Comparing Eqs.(\ref{Fexp},\ref{Fmodel}) we obtain density of {\it independent} 
excitations 
\be
d_{\rm exc}~=~c~d~.
\ee 
We can conclude that $c=d_{\rm exc}/d$ measures correlations between kinks: 
$c<1$ means bunching and an eventual $c>1$ would mean anti-bunching of kinks 
randomly distributed along the spin chain.

This simple interpretation of $c$ follows from the fact that any ground state 
$|0_{g,\Gamma}\rangle$ is a BCS state of kinks $\gamma^{(0)}$, see Eq. (\ref{expZ})
where $Z_{ab}$ is a wave-function for a (Cooper) pair of kinks. Depending on $g$ and 
$\sigma$, this BCS state can be either a condensate of tightly bound Cooper pairs of 
kinks (with $c=1/2$), or just a weakly coupled BCS state with pairing 
(anti-)correlations manifesting as the (anti-)bunching of kinks. 

The BCS state is particularly simple at weak magnetic fields $g_n=g+\Gamma_n$ when,
crudely speaking, the ground state $|0_{g,\Gamma}\rangle$ in Eq. (\ref{0pert}) is 
approximately the ferromagnet $|\uparrow_1\uparrow_2...\uparrow_N\rangle$ but with 
occasional spins reversed to $|\downarrow_n\rangle$ by weak magnetic fields $g_n\sigma_n^x$. 
Each reversed spin is a tightly bound Cooper pair of two kinks sitting on nearest neighbor 
bonds with a relative distance of $r=1$ lattice sites between the two kinks. No wonder
that $d_{\rm exc}=d/2$ in this case.

As we could see in Fig. \ref{FigStatic}C, the picture with tightly bound Cooper 
pairs is accurate for weak disorder in the ferromagnetic phase, but with $g$ increasing 
above $g_c$ the Cooper pairs begin to dissociate into free kinks and antikinks 
and $c$ increases well above $\frac12$. This dissociation can be clearly seen
in Fig. \ref{FigStatic}D where we show a correlation function $C_r$ 
\be
C_r~\sim~\sum_m~\left|Z_{m+r,m}\right|^2~
\label{C_r}
\ee
between two kinks making a Cooper pair. Here $Z_{a,b}$ is the unnormalized
wavefunction of a Cooper pair of two kinks in the BCS state (\ref{expZ}) and 
$r$ is a relative distance between the two kinks. The $C_r$ shown in 
Fig. \ref{FigStatic}D is localized when $g<=g_c$ (tight Cooper pairs) and 
delocalized when $g>g_c$ (dissociated pairs). 

A similar $C_r$ is shown in Fig. \ref{FigStatic}E for a finite $\sigma=0.8$ 
in the ferromagnetic phase at $g=0$. In addition, the same figure shows 
reduced probability distribution
\be
P_n~\sim~\sum_m~\left|Z_{m,n}\right|^2
\label{P_n}
\ee
for a single kink. Unlike in the pure case, this $P_n$ is not uniform and shows 
Anderson localization.
  
In the ferromagnetic phase below $g_c$ all kinks are bound into tight Cooper pairs. 
The small density of Cooper pairs, each of them a tightly bound pair of kink and 
antikink, does not destroy long range ferromagnetic order in $\sigma^z$. With $g$ 
increasing above $g_c$ the transition to the paramagnetic phase takes place when 
the Cooper pairs begin to dissociate into free kinks and antikinks which scramble 
long-range ferromagnetic correlations. 

In the context of adiabatic quantum computation there are two generally accepted
measures of how far the final state $\rho_S(0)$ from the desired final ground 
state $|0\rangle$ is. One is energy of excitation of the final state above the energy 
of the desired final ground state, and the other is fidelity between the final state 
and the final ground state. Both quantities are tractable in our model, where 
the excitation energy is proportional to the number of kinks. We study them in the
following two Sections.

\section{Density of kinks after a quench}

In our simulations of a quench the system is initially prepared in its ground state 
at a large initial value of $g\gg1$ i.e. in a Bogoliubov vacuum state for quasiparticles 
at an initial $g\gg 1$. As the magnetic field is being turned off to 
zero across $g_c$, the state of the system $|\psi(t)\rangle$ is getting excited from 
its adiabatic ground state. However, in a similar way as in 
Refs. \cite{Dziarmaga2005,JDrandom}, we use the Heisenberg picture where the state 
remains a vacuum for quasiparticle operators
\be
\tilde\gamma_m=u_{nm}^*(t)c_n+v_{nm}^*(t)c_n^\dagger~
\label{tildegamma}
\ee 
with the Bogoliubov modes $u_{nm}(t)$ and $v_{nm}(t)$ solving time-dependent 
Bogoliubov-de Gennes equations
\bea
i\frac{du_{n,m}^\pm}{dt} = 2 g_n(t) u_{n,m}^\mp - 2 u_{n-1,m}^\mp ~.
\label{tBdG}
\eea
Initially each mode is a positive frequency eigenmode of the stationary Eqs.(\ref{BdG}). 
Equations (\ref{tBdG}) were integrated by a split-step method for each realization of 
$\Gamma_n$. Their solutions $u^\pm_{nm}(0)$ at the final $g=0$ determine final 
states $|\psi(0)\rangle$ whose average over disorder gives final density matrix 
$\rho_s(0)=\overline{|\psi(0)\rangle\langle\psi(0)|}$. For each realization of $\Gamma_n$,
a final state $|\psi(0)\rangle$ is a vacuum for $\tilde{\gamma}_m$'s which 
are related to the kinks $\gamma^{(0)}_m$ by the transformation 
Eq. (\ref{UV}). Final densities of kinks follow from Eq. (\ref{dVV}).

\begin{figure}
\includegraphics[width=0.999\columnwidth,clip=true]{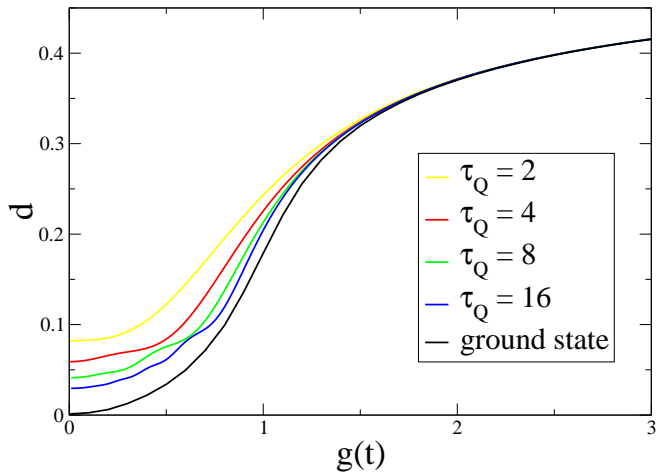}
\caption{ Density of kinks during a quench as a function of $g(t)$. Here each plot
is a single realization with $\sigma=0.1$ on a lattice of $N=512$ sites. For the early 
large $g$ the density follows the density $d(g,\sigma)$ in a ground state 
$|0_{g,\Gamma}\rangle$, but as the quench is approaching the critical point $g_c\approx 1$ 
the evolution becomes non-adiabatic and the density gets excited above its ground state 
level. The slower is the quench, the closer to the critical point the non-adiabatic 
stage begins and the quench is probing the critical point more closely. 
}
\label{Figd(t)}
\end{figure}

For the system Hamiltonian (\ref{HS}) at the final $g=0$ the final excitation 
energy is simply twice the final number of kinks excited in the desired ferromagnetic 
ground state. In Figure \ref{Figd(t)} we show density of kinks $d$ as a 
function of $g(t)$. For the early large $g$ the density follows the density 
$d(g,\sigma)$ in a ground state $|0_{g,\Gamma}\rangle$, which is also shown in 
Fig. \ref{FigStatic} B, but as the quench is approaching the critical point 
$g_c\approx 1$ the evolution becomes non-adiabatic and the density $d$ is 
excited above its ground state level $d(g,\sigma)$. The slower is the quench, 
the closer to the critical point the non-adiabatic stage begins and the quench 
is probing the critical point more closely.

\begin{figure}
\includegraphics[width=0.999\columnwidth,clip=true]{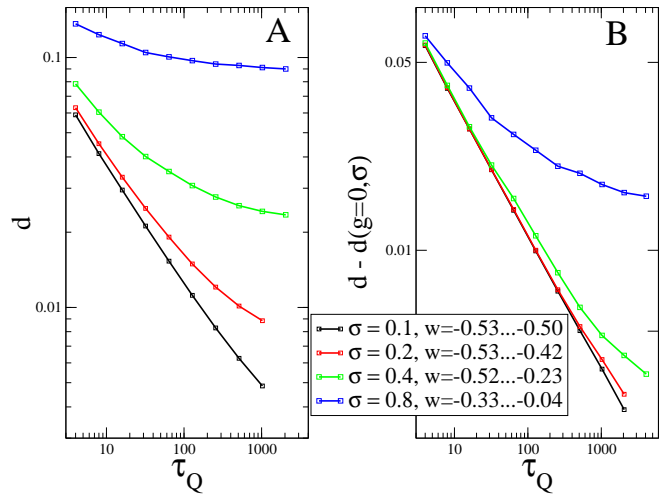}
\caption{ 
In A final density of kinks at $g=0$ is shown as a function of $\tau_Q$ on a
$N=512$ chain. For large $\tau_Q$ the density tends to saturate at the average density 
of kinks $d_{g=0,\sigma}$ in the ground state of the random Hamiltonian (\ref{Hsigma}) 
shown in Fig. (\ref{FigStatic})B. In B we show difference $\delta d=d-d_{g=0,\sigma}$ 
which is density of kinks excited above the ground state of the random Hamiltonian
at $g=0$. This difference can be attributed to the non-adiabaticity of the transition 
described by KZM. 
}
\label{Figd}
\end{figure}

The final density of kinks at $g=0$ is shown in Fig. \ref{Figd}A. For large $\tau_Q$, 
the density tends to saturate at $d(g=0,\sigma)$ shown in Fig. (\ref{FigStatic})B 
i.e. the density of kinks in the ground state of the random Hamiltonian 
(\ref{Hsigma}). In Fig. \ref{Figd}B we show a difference $\delta d=d-d(g=0,\sigma)$ 
which can be attributed to the non-adiabaticity of the transition described by KZM. 
If $\delta d=\alpha\tau_Q^w$, then in the log-log plot of Fig. \ref{Figd}B we would 
see a line $\log_{10}\delta d=\log_{10}\alpha+w(\log_{10}\tau_Q)$, but this is not 
the case when $\tau_Q\gg\sigma^{-2}$ as in Eq. (\ref{strong}). At best we can think 
of a local slope $w(\tau_Q)$ which can be estimated by fitting to pairs of nearest 
neighbor data points. In the legend we give ranges of local slopes $w$ obtained for 
different $\sigma$'s (with error bars on their last digits). For weak $\sigma$ and 
small $\tau_Q$ the slope $w$ is close to the $-1/2$ characteristic for the pure model 
(\ref{HS}): fast quenches, when $\tau_Q\ll\sigma^{-2}$, become non-adiabatic far 
enough from the critical point not to see any effect of weak disorder. At stronger 
$\sigma$ or longer $\tau_Q$ the local slopes are less steep and for a fixed $\sigma$ 
they become less steep with increasing $\tau_Q$. For example, at the strongest 
$\sigma=0.8$ the local slope falls to a mere $|w|=0.04$ for the longest $\tau_Q$. These 
observations are consistent with the predicted logarithmic dependence of the dynamical 
correlation length $\hat\xi$ in Eq. (\ref{xirandom}).

\section{ Fidelity and correlations after a quench }

In the context of adiabatic quantum computation, it is important to know how close 
the final state $\rho_S(0)$ to the desired ground state of the final Hamiltonian 
$H_S$ is. The closeness is measured by fidelity 
\be
F~=~
\langle 0|\rho_S(0)|0\rangle~=~
\overline{
\langle 0|\psi(0)\rangle\langle\psi(0)|0 \rangle
}~
\label{F}
\ee 
given by Eq. (\ref{expTrlog}). 

Without decoherence, or in a pure Ising chain with $\sigma=0$, the fidelity can be 
obtained analytically from the exact solution in Refs. \cite{Dziarmaga2005,Cincio}. 
$F$ is a probability that not a single pair of $\gamma^{(0)}$-quasiparticles with 
opposite quasimomenta $(k,-k)$ is excited after a quench,
$
F=\prod_{k>0}(1-p_k).
$
Here $p_k\approx\exp(-2\pi\tau_Q k^2)$ when $\tau_Q\gg1$. When $Nd\gg 1$ we obtain
\bea
\ln F \approx
-\frac{Nd}{2}
\frac{  \int_0^\infty ds~ \ln\left[1-e^{-s^2}\right]  }
     {  \int_0^\infty ds~ e^{-s^2}                    } \approx
-1.3Nd~.
\label{Fsigma0}
\eea 
Here $d=\int_{-\pi}^\pi\frac{dk}{2\pi}p_k=1/2\pi\sqrt{2\tau_Q}$. $F$ is exponential
in $N$ as in the static case Eq. (\ref{Fexp}). Given that $d\ll 1$ for $\tau_Q\gg 1$
we obtain that $c\approx1.3>1$ implying anti-bunching of kinks.

\begin{figure}
\includegraphics[width=0.999\columnwidth,clip=true]{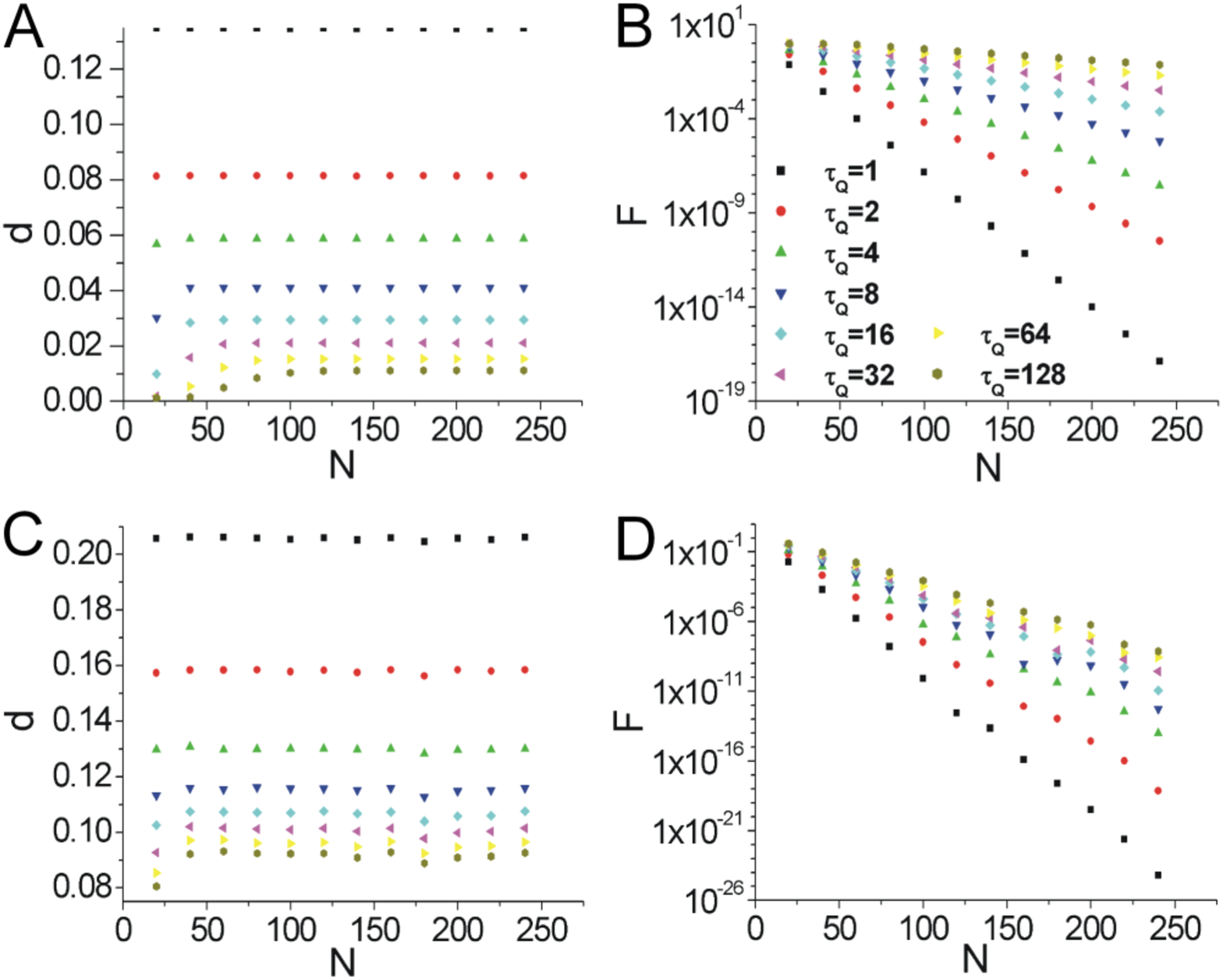}
\caption{ 
In A and C final kink density $d$ for different $\tau_Q$ as functions of 
lattice size $N$ for $\sigma=0.1$ (panel A) and $\sigma=0.8$ (panel C). 
In B and D fidelity $F$ for $\sigma=0.1$ (panel B) and $\sigma=0.8$ (panel D).
With inreasing $N$ the density saturates at an asymptotic $d(\tau_Q,\sigma)$
when $N\gg1/d(\tau_Q,\sigma)$. In the same asymptotic regime the fidelity
becomes exponential in $N$. Here the averages are taken over $N_R$ realizations 
with $N_RN\geq2048$.
}
\label{dF}
\end{figure}
\begin{figure}
\includegraphics[width=0.999\columnwidth,clip=true]{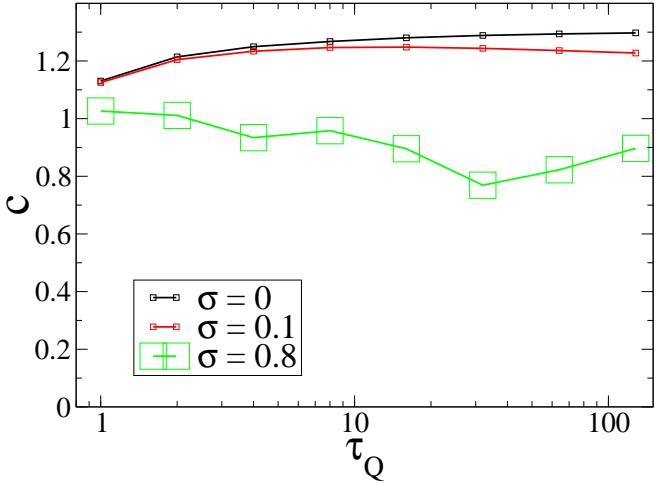}
\caption{ 
Correlation coefficients $c$ for different $\tau_Q$ and $\sigma$ obtained by 
linear fits to the exponential tails of fidelity in Figs. \ref{dF}B and D. 
}
\label{cfig}
\end{figure}

\begin{figure}
\includegraphics[width=0.999\columnwidth,clip=true]{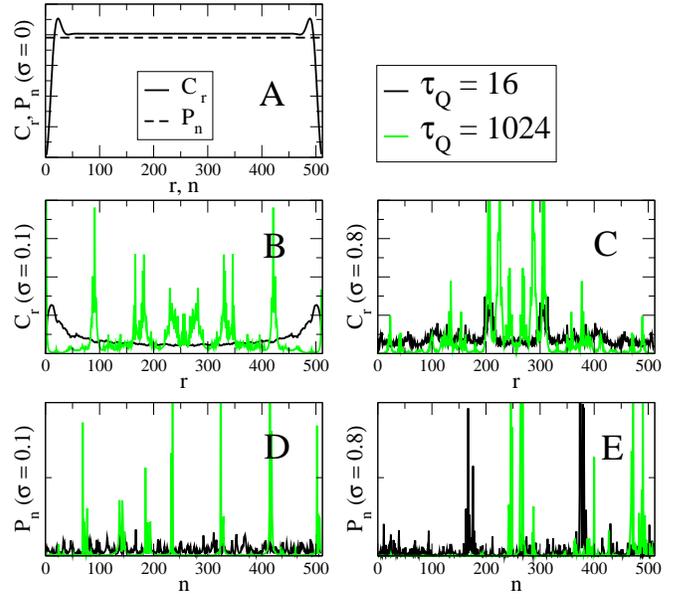}
\caption{ In A a correlation function $C_r$ between two kinks in a Cooper 
pair and probability distribution $P_n$ for a kink on a $N=512$ periodic lattice  
after a quench with $\sigma=0$ and $\tau_Q=16$.
In B, C, D, and E we show typical final $C_r$ and $P_n$ after quenches 
with $\sigma=0.1,0.8$ and $\tau_Q=16,1024$. Here all $C_r$ and $P_n$
for a given $\sigma$ come from the same realization of disorder $\Gamma_n$.
Both $C_r$ and $P_n$ show Anderson localization. When $\tau_Q$ or $\sigma$
are large, then $C_r$ is localized around those $r$'s which can be 
identified as distances (modulo periodic boundary conditions) between the 
localization centers in the corresponding $P_n$, with the exception of 
$r\approx 0$ which is avoided because two fermionic kinks do not like to 
choose the same localization center. 
}
\label{CCCC}
\end{figure}

The anti-bunching can also be seen in the state $|\psi(0)\rangle$ after a quench
which is a BCS state of kinks in Eq. (\ref{expZ}). Figure \ref{CCCC}A shows 
a probability distribution $P_n$ for a kink in Eq. (\ref{P_n}) and the correlation 
function $C_r$ between two kinks in a Cooper pair in Eq. (\ref{C_r}) after a quench 
with $\sigma=0$ and $\tau_Q=16$. We have $C_0=0$ because the kinks are fermions, there 
is a broad maximum in $C_r$ in the range $|r|=20\dots40$, and a flat distribution for 
$|r|>40$ when the Cooper pair is dissociated. If $C_r$ were flat everywhere (except $r=0$), 
then $c=1$, but the broad maximum means that even when the kinks get close to each 
other they prefer to keep a safe distance in the range $20\dots 40$. This short 
range repulsion is consistent with the anti-bunching observed in Eq. (\ref{Fsigma0}) 
where $c=1.3>1$.

The same anti-bunching can also be seen in the ferromagnetic correlation
function after a quench at $g=0$
\be
\langle \sigma_{i}^z \sigma_{i+R}^z \rangle \simeq
\exp\left( -1.55Rd \right)
\cos\left(  2.95Rd - \varphi \right)
\label{crystal}
\ee
accurate when $1\ll R\ll \sqrt{\tau_Q}\log\tau_Q$, see Ref. \cite{Cincio}. The 
oscillatory cosine factor means that kinks tend to order into a crystal lattice.
This very imperfect crystalline order implies the anti-bunching seen in $c=1.3>1$. 
We can conclude that in the state after a quench with $\sigma=0$ there is a similar 
connection between $c$, $C_r$, and ferromagnetic correlations as in the static case.

\begin{figure}
\includegraphics[width=0.999\columnwidth,clip=true]{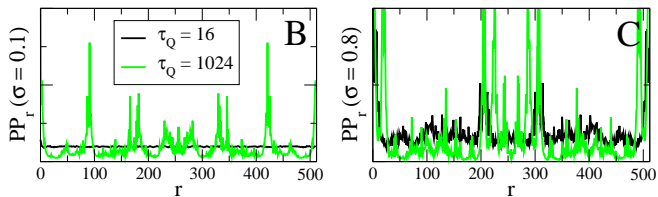}
\caption{ In B and C convolutions in Eq. (\ref{PP}) corresponding to $C_r$'s in 
Figs. \ref{CCCC}B and D obtained from distributions $P_n$ in Figs. \ref{CCCC}D and E.
Except for $\tau_Q=16,\sigma=0.1$ (black in panels B), the convolutions are very close 
to their corresponding $C_r$'s everywhere apart from $r$ close to $0$ (and $N$).    
}
\label{PP}
\end{figure}

For quenches with $\sigma>0$ we also find an exponential tail
\be
F(\tau_Q,\sigma) \sim (1-cd)^N
\ee 
when $F\ll 1$, compare panels B and D in Fig. \ref{dF}. Here $d(\tau_Q,\sigma)$ is 
an asymptotic value of average kink density obtained for a sufficiently large lattice
size $N$, such that $Nd(\tau_Q,\sigma)\gg1$, see panels A and C in the same 
Fig. \ref{dF}. In contrast, when $N\ll 1/d(\tau_Q,\sigma)$, then a finite gap at $g=g_c$
results in an adiabatic transition and a final density of kinks which is less than the 
asymptotic $d(\tau_Q,\sigma)$ for large $N$. 

Fits to the exponential tails of fidelity in Figs. \ref{dF}B and D give correlation 
coefficients $c$ for different $\tau_Q$ and $\sigma$ which are shown in Fig. \ref{cfig}. 
For comparison, we also show in the same figure $c$ at $\sigma=0$ which saturates 
at $c=1.3$ when $\tau_Q\gg 1$. The results suggest that for a strong $\sigma$ or large 
$\tau_Q$, when even a weak $\sigma$ has strong effect, the correlaton coefficient decays 
towards $c\approx 1$ suggesting random Poissonian trains of kinks and exponential 
ferromagnetic correlation functions. 

This picture is corroborated by typical correlation functions $C_r$ and probability 
distributions $P_n$ shown in panels B, C, D, and E of Fig. \ref{CCCC}. Here it is clear, 
especially for the larger $\tau_Q=1024$ (green) or the stronger $\sigma=0.8$ (panels C 
and E), that a kink gets localized in isolated Anderson localization centers. 
A close inspection of corresponding $C_r$'s reveals that $C_r$ is localized at $r$'s 
equal to distances between the localization centers in $P_n$. There is only one exception
from this rule: $C_r$ is negligible when $r\approx 0$ or $N$ because, apparetly, 
a (Cooper) pair of fermionic kinks avoids being trapped in the same localization 
center. It seems that each kink in a Cooper pair chooses its localization center 
at random, independently of the other kink, except for avoiding the same localization
center. A quantitative proof of this simple picture is provided in 
Figs. \ref{PP}B and C where we plot a convolution
\be
PP_r~=~\sum_{n=1}^N P_n~P_{n+r}~.
\label{PPr}
\ee
If the two kinks in a Cooper pair were independent, then this convolution would 
reconstruct the corresponding $C_r$'s in Figs. \ref{CCCC}B and C. Comparing panels 
B and C in the two figures we see that it does, with the expected exceptions
when $r\approx 0,N$ and in the case of a fast quench ($\tau_Q=16$) at weak $\sigma=0.1$ 
(black in Fig. \ref{CCCC}B). We can conlude that, with some idealization, for long $\tau_Q$ 
or strong $\sigma$, final kinks are distributed as if each kink were choosing at random 
one of the isolated Anderson localization centers randomly distributed along 
the chain. This approximate Poissonian model is consistent with the observed
$c\approx1$. 

When $\tau_Q\sigma^2\gg 1$, then the final states at $g=0$ are 
qualitatively different from the ground state of the random Hamiltonian (\ref{Hsigma})
at $g=0$: the (green) fragmented $C_r$'s in Figs. \ref{CCCC}B and C are qualitatively 
different from the (black) $C_r$ in Fig. \ref{FigStatic}E localized around $r=0$. 
In other words, all kinks in the ground state, of density $d(g=0,\sigma)$, are tightly 
bound into Cooper pairs which do not destroy ferromagnetic long range order, while all 
kinks in a final state, of density $d$, contribute to exponentially decaying ferromagnetic 
correlations even though for slow quenches the difference $\delta d=d-d(0,\sigma)$,
whose origin can be attributed to non-adiabaticity, is small as compared to $d(0,\sigma)$.    
 
\begin{figure}
\includegraphics[width=0.999\columnwidth,clip=true]{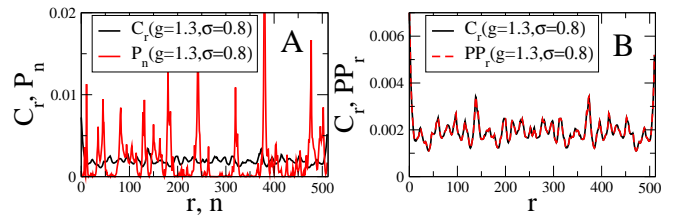}
\caption{ In A $C_r$ and $P_n$ in a ground state of the random Hamiltonian 
(\ref{Hsigma}) at $\sigma=0.8$ and $g=1.3$ just above $g_c$. Here $P_n$ shows
similar fragmentation into isolated Anderson localization centers as in the final
states after a quench with long $\tau_Q$ or large $\sigma$ in Figs. \ref{CCCC}.
In B we show both the $C_r$ from panel A and a convolution $PP_n$ in Eq. (\ref{PPr})
obtained form the $P_n$ in panel A. The two plots are identical, except for $r=0$, 
demonstrating independence of kinks.
}
\label{PPhatg}
\end{figure}

These striking properties of the final states after a quench are inherited from
the ground state of the random Hamiltonian (\ref{Hsigma}) just above $g_c$. Its 
properties are relevant here because, as we could see in Fig. \ref{Figd(t)}, in a 
slow quench a state $|\psi(t)\rangle$ follows an adiabatic ground state of the 
random Hamiltonian until a $\hat g$ just above $g_c$ when, in accordance with KZM, 
the evolution becomes non-adiabatic. In Fig. \ref{PPhatg}, we show some properties 
of the ground state just above $g_c$ where $P_n$ is fragmented into isolated 
localization centers, and $C_r$ is identical with $PP_r$ everywhere except $r=0$. 
Apparently, the following evolution from $\hat g$ to $g=0$ does not change this 
qualitative picture.

\section*{ Conclusion }

A static spin environment increases non-adiabaticity of the transition in a dramatic 
way: density of quasiparticles (kinks) decays no longer as a power of the transition 
time but in a much slower logarithmic way. This means, in the context of adiabatic 
quantum computation, that coupling to a static environment may transform a polynomial 
computational problem into a non-polynomial one. 

Fidelity between a final state after a quench and the desired final ground state 
decays exponentially with a chain size. The rate of this decay is equal to the 
density of kinks times a correlation coefficient equal, for fast transitions
and weak decoherence, to $1.3$ and, for slow transitions or strong decoherence,  
close to $1$. Corresponding kink-kink correlations are, respectively, anti-bunching 
in a near-crystalline ordering and a simple Poissonian distribution of kinks in 
isolated Anderson localization centers randomly distributed along a chain.

\section*{ Acknowledgements } 

We would like to thank Wojciech Zurek for discussions. Work of L.C., J.D., J.M. 
and M.R. was supported in part by Polish government scientific funds (2008-2011) 
as research projects N N202 175935, N N202 079135, N 202 059 31/3195, and 
N N202 174335 respectively, and in part by Marie Curie Actions Transfer of Knowledge 
project COCOS (contract MTKD-CT-2004-517186).


\end{document}